\documentclass[12pt]{article}
\usepackage{graphicx}

\begin{document}

\parindent=1.7cm
\begin{center}
 {\bf \Large Crystal structure of mixed fluorites
 $Ca_{1-x}Sr_xF_2$ and $Sr_{1-x}Ba_xF_2$ and luminescence of $Eu^{2+}$ in this crystals}.
\end{center}
 \vspace{2ex}
\begin{center}
 A.E.~Nikiforov,  A.Yu.~ Zaharov,V.A.~Chernyshev(Vladimir.Chernyshev@usu.ru),
 M.Yu.~Ougrumov, S.V.~Kotomanov
\end{center}
 \vspace{2ex}
\begin{center}
 Urals State University,Ekaterinburg,Russia
\end{center}
 \vspace{2ex}

 {\bf Abstract.}
 Within the framework of the virtual crystal method
implemented in the shell model and pair potential approximation
the crystal structure of mixed fluorites $Ca_{1-x}Sr_xF_2$ and
$Sr_{1-x}Ba_xF_2$ has been calculated. The structure of impurity
center $Eu^{2+}$ and the distance $E^{2+}-F^-$ in this  crystals
have been also calculated. The low level position of excited
$4f^65d$ configuration of the $Eu^{2+}$ ion has been expressed
using phenomenological dependence on distance $Eu^{2+}-F^-$. The
dependences of Stokes shift and Huang-Rhys factor on $x$ have been
received for yellow luminescence in $Sr_{1-x}Ba_xF_2:Eu^{2+}$. The
value $x$, for which the $e_g$ -level of $Eu^{2+}$ ion will be in
conduction band in $Sr_{1-x}Ba_xF_2:Eu^{2+}$ has been calculated.

\section{Introduction}

 Fluorites {\it $CaF_2$ , $SrF_2$ , $BaF_2$} and mixed crystals on their base
  attract attention of researchers more than four decades \cite{An,CaSr,SrBa,Fefl,KapFefl}.\
  The crystals, doped by rare--earth ions, are good luminophors and are  a basis for solid state lasers.
  The  properties of luminescence and absorption depend on  electronic structure of matrix crystal.
  The blue luminescence with zero phonon line (ZPL) is observed in $Ca_{1-x}Sr_xF_2:Eu^{2+}$. The yellow luminescence without ZPL is observed in
 $Sr_{1-x}Ba_xF_2:Eu^{2+}$ at $x>0{,}2$.

\section{Calculation of the crystal structure \protect\\$Ca_{1-x}Sr_xF_2$ and $Sr_{1-x}Ba_xF_2$.}

The method of virtual crystal have been implemented within the
framework of the pair potential approximation and shell model. The
method used has been described in previous work \cite{NikUgr} . We
have calculated the crystal structure of mixed fluorites
$Ca_{1-x}Sr_xF_2$ and $Sr_{1-x}Ba_xF_2$ and of impurity center
$Eu^{2+}$  in this crystals. Dependence of a lattice constant of
$Sr_{1-x}Ba_xF_2$ from $x$ is given on fig.1.

\begin{figure}
 \includegraphics[height=1.2\linewidth, width=\linewidth,clip=true]{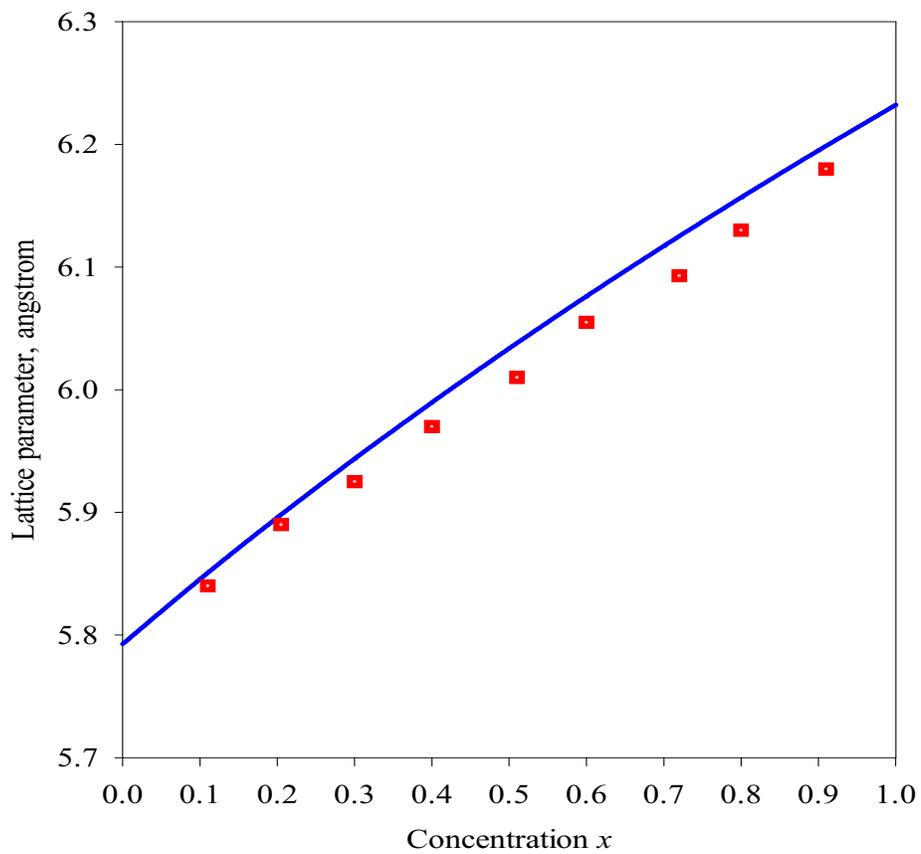}
 \caption{The lattice constant of
$Sr_{1-x}Ba_xF_2$ from $x$. Square symbol refer to experiment
\protect\cite{An}.}
\end{figure}
The calculated dependence is in agreement with Wegard rule, though
at medial concentrations is appreciable some difference from
results of calculation. Similar dependence has been calculated for
$Ca_{1-x}Sr_xF_2$.
\section{Impurity ion $Eu^{2+}$ in crystals $Ca_{1-x}Sr_xF_2$ and $Sr_{1-x}Ba_xF_2$}
The luminescence and adsorption spectra of $Eu^{2+}$ in
$Ca_{1-x}Sr_xF_2$ , $Sr_{1-x}Ba_xF_2$ deals with
interconfigurational transitions between low excited levels of
$4f^65d$ configuration and by $^8S(4f^7)$ ground state
\cite{KapFefl}. The ion $Eu^{2+}$ is in center of cube formed by
eight fluorines $F^-$. The splitting of $^8S(4f^7)$ ground state
is small in cubic crystal field and does not exceed $0{,}2cm^{-1}$
\cite{Ryt}. Level $5d$ is enough feebly bound by with $4f^6$
orbitals in excited configuration $4f^65d$. The level splits to
$e_g$ and $t_{2g}$ levels $12-16\times10^3cm^{-1}$ \cite{KapFefl}.
The split considerably exceeds $LS$ interaction in $t_{2g}$ state
and multiplet split of $4f^6$ levels \cite{Mtsl}. Spectrum of the
impurity ion is substantially defined by distance impurity
ion--ligand. We have calculated the distance $Eu^{2+}-F^-$ in the
crystals and then have calculated a phenomenological dependence of
standing $e_g$ and $t_{2g}$ levels from the distance. Standing of
$e_g$ level in $Ca_{1-x}Sr_xF_2:Eu^{2+}$, is expressed by
dependence:
\begin{equation}
 \label{one}
 \nu(r)=C+\frac{A}{r^n}-\frac{B}{r^k}
\end{equation}
where $n=12$, $k=5$. The first term determines standing of $5d$
level in free ion $Eu^{2+}$,
\begin{figure}
\includegraphics[height=1.2\linewidth, width=\linewidth,clip=true]{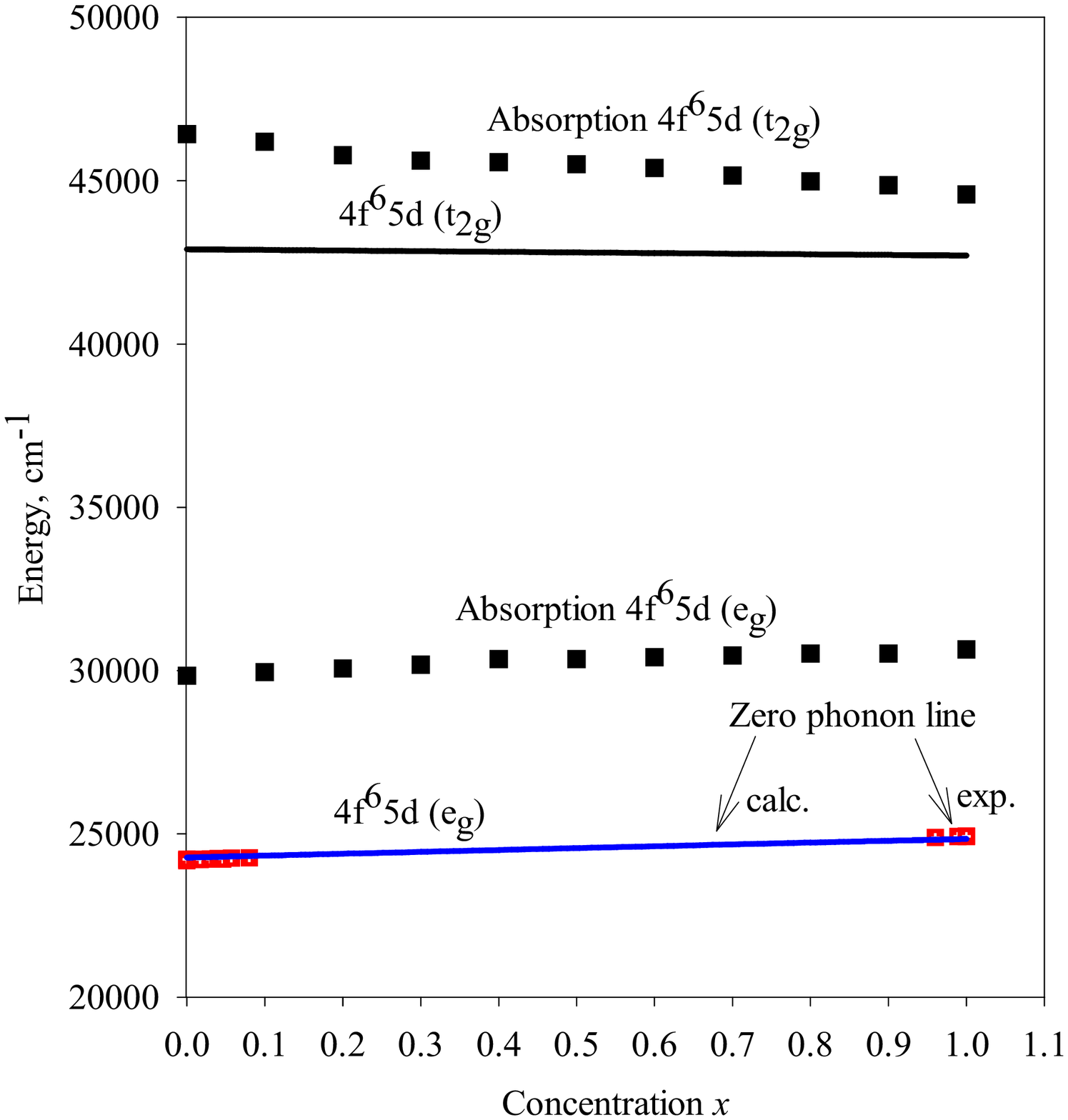}
 \caption{$t_{2g}$ and $e_g$ levels of $Eu^{2+}$.
 Square symbol refer to experiment\protect\cite{CaSr}.}
\end{figure}
the second term determines shift of the level at placing $Eu^{2+}$
in a crystal and third term deals with influence of crystal field
on $t_{2g}-e_{g}$ splitting. Parameters $A$,$B$,$C$ have been
received by fitting the dependence to standing ZPL in $CaF_2$ and
$SrF_2$ crystals \cite{CaSr} and  $t_{2g}-e_g$ splitting in $MeF_2
(Me=Ca,Sr,Ba)$ \cite{SrBa,Fefl}. The parameters are: $A=439{,}7
\times10^6 cm^{-1} \times \AA^{12}$ , $B=280 \times10^5cm^{-1}
\times \AA^5$, $C=36940 cm^{-1}$. The dependence of  ZPL position
from $x$ we can receive using calculated distance $Eu^{2+}-F^-$ in
$Ca_{1-x}Sr_xF_2:Eu^{2+}$ at various $x$. It is possible to
calculate the $t_{2g}$ -level position by means of \cite{Ryt}
taking into account that third term in \cite{Ryt} is equal to
$6Dq$ (fig.2). The difference of the calculated $t_{2g}$-level
position with  short-wave adsorption peak can be explained by
Stokes shift in adsorption.

The blue and yellow luminescence is observed in
$Sr_{1-x}Ba_xF_2:Eu^{2+}$ for $x\in {[0{.}2,0{.}5]}$ in intervals
$430-450$ and $500-580$ nm. accordingly. The yellow luminescence
at $x>0.5$ is only observed \cite{SrBa}. Yellow luminescence due
to interconfigurational transitions between levels of impurity
exciton (which is formed by transition of electron to twelve
nearby cations) and the ground state $^8S(4f^7)$
\cite{SrBa,MClure} (fig.3).
\begin{figure}
\includegraphics[height=1.2\linewidth, width=\linewidth,clip=true]{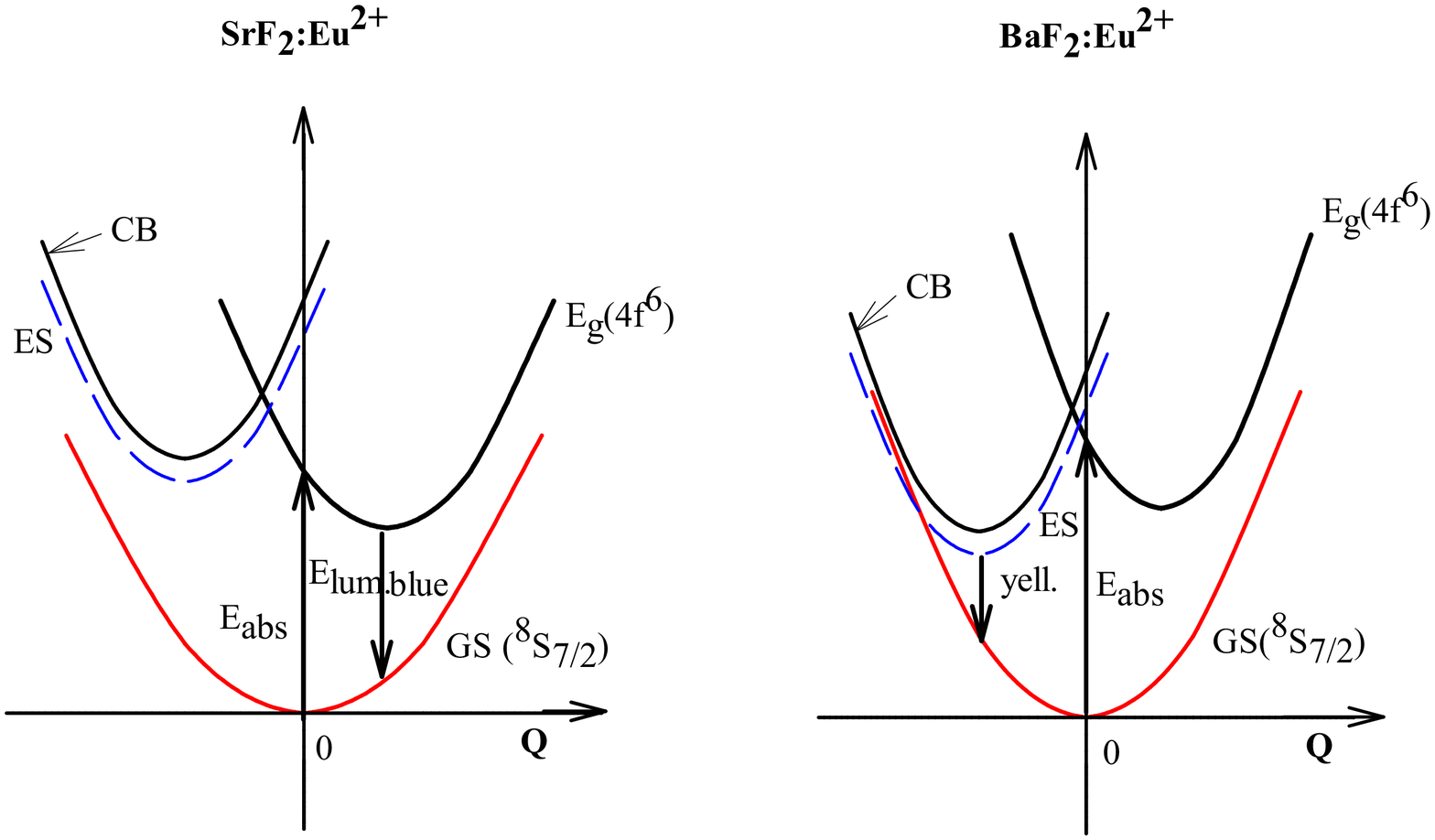}
\caption{Configuration coordinate diagrams for describing of the
blue and yellow luminescence in $CaF_2:Eu^{2+}$ and
$SrF_2:Eu^{2+}$ \cite{MClure}}
\end{figure}
Position of $e_g$-level can be calculated in
$Sr_{1-x}Ba_xF_2:Eu^{2+}$ using the expression ~(\ref{one}). The
top of valence band is formed by $2p$-states of fluorine in
$SrF_2$ and $BaF_2$. The bottom of conduction band is formed by
$s$-states of cation. Forbidden zone varies practically linearly
in row $CaF_2$, $SrF_2$, $BaF_2$ \cite{Rubl}. The distance from
$e_g$ level to bottom of conduction band in $SrF_2:Eu^{2+}$ was
taken from  McClure work \cite{MClure}. Assuming the position of
$2p$-states of fluorine does not change at replacing cations
$Sr^{2+}$ by $Ba^{2+}$, it is possible to calculate the changing
of conduction zone bottom as a function from $x$. According to
ours calculation, $e_g$-level will be in a conduction band for
$x=0{.}2$ in $Sr_{1-x}Ba_xF_2:Eu^{2+}$. The yellow luminescence
will begin at this $x$ in $Sr_{1-x}Ba_xF_2:Eu^{2+}$ \cite{SrBa}.We
have calculated configuration curves  connected with $A_{1g}$
-coordinate knowing how the energy of the crystal $E$ depends on
compression or expansion of a fluorine cube  which surrounds
$Eu_{2+}$. The deduced dependences  $E(Q)$ (where $Q$-is the
symmetrical coordinate) for $Sr_{1-x}Ba_xF_2:Eu_{2+}$ are close to
parabolic form $kQ^2/2$. Similar calculations was carried out for
exciton state. The coefficient $k$ dependences  of the
configurational curves on $x$ are received:
$k_{ES}=33{.}95-5{.}26*x$,\ $k_{GS}=21{.}18-4{.}61*x$,\ where
$x\in{[0,1]}$, ( in $eV \times{\AA^{-2}}$). The coefficients
$k_ES$ and $k_GS$ are decreasing at increasing the barium
concentration. It compounded with the fact that the elastic
modules $BaF_2$ are less than $SrF_2$. The fluorine cube is
exposed to compression at formation of the exciton, the
compression depends linearly on the concentration $x$ in
$Sr_{1-x}Ba_xF_2:Eu^{2+}$: $\triangle R=0{.}143+0{.}036*x$ (in
$\AA$). We can receive the Stoks shift for yellow luminescence as
we know  $k_GS$   and change of distance $Eu^{2+}-F^-$ at
transition from exciton to the ground state: $E_s=1090*x+5011$ (in
$cm^{-1}$). The yellow luminescence have been observed in
$Sr_{1-x}Ba_xF_2:Eu^{2+}$ at $\in{[0{.}2,1]}$, and with increasing
$x$ its peak is shifting to the long-wave part of the spectrum
\cite{SrBa}. According to our calculation the Stoks shift is
increased on $800cm^{-1}$. We can estimate $A_{1g}$ frequency of a
cube from eight ions $F^-$:
\begin{equation}
 \label{two}
  \nu_{A_1g}=\sqrt{\frac{k_{GS}}{m_F}}
\end{equation}
where $m_F$ --a mass of the fluorine. According to our
calculations $n=547-63*x$, (n in $cm^{-1}$). We can calculate
Huang-Rhys factor for yellow luminescence as we know $E_s$ and
$n$. The factor is increasing from 9 to 12 at increasing $x$ from
$0$ to $1$ in $Sr_{1-x}Ba_xF_2:Eu^{2+}$ .
\section{Conclusion}
The method of virtual crystal implemented in shell model and
framework of pair potential approximation allows us to calculate
the crystal structure and lattice constant for mixed fluorites
$Ca_{1-x}Sr_xF_2$ and $Sr_{1-x}Ba_xF_2$. This method allows us to
calculate the distance $Eu^{2+}-F^-$ in doped crystals\linebreak
$Ca_{1-x}Sr_xF_2:~Eu^{2+}$ and $Sr_{1-x}Ba_xF_2:~Eu^{2+}$. The low
level position of excited $4f^65d$ configuration of an ion
$Eu^{2+}$ is expressed by phenomenological dependence on
$Eu^{2+}-F^-$ distance . The dependences of Stokes shift and
Huang-Rhys factor on $x$  have been calculated for yellow
luminescence in $Sr_{1-x}Ba_xF_2:Eu^{2+}$. According to our
calculation the Stokes shift is increasing by $1000cm^{-1}$, the
Huang-Rhys factor is increasing from $9$ to $12$ while $x$ changes
from $0$ to $1$. According to calculation the  $e_g$ level of
$Eu^{2+}$ ion will be within the  conduction band for $x\ge0{.}2$
in $Sr_{1-x}Ba_xF_2:Eu^{2+}$. \vspace{2ex}

{\small This study was supported by grant no. $REC 005 (CRDF)$.}

\end{document}